\documentstyle[aps,multicol,epsfig,eqsecnum]{revtex}
\def\be{\begin{eqnarray}}
\def\ee{\end{eqnarray}}
\newcommand{\ba}{\begin{array}}
\newcommand{\ea}{\end{array}}

\def\l{\langle}
\def\r{\rangle}

\begin{document}
\title{ 
Equally-distant partially-entangled  alphabet
states for quantum channels
}
\author{
M\'ario Ziman$^{1}$ and 
Vladim\'{\i}r Bu\v{z}ek$^{2}$\thanks{
{\bf On leave from}: 
the Institute of Physics, Slovak Academy of Sciences, D\'ubravsk\'a cesta 9,
842 28 Bratislava, Slovakia, and Faculty of Informatics, Masaryk University,
Botanick\'a 68a, 602 00 Brno, Czech Republic
}
}

\address{
$^{1}$
Faculty of Mathematics and Physics, Comenius University,
Mlynsk\'a dolina F2, Bratislava 842 15, Slovakia\\ 
$^{2}$ SOKEN-DAI, The Graduate University for Advanced Studies, Shonan
Village,
Hayama, Kanagawa 240-0193, Japan \\ 
}

\date{1 June 2000}
\maketitle
\begin{abstract}
Each 
Bell state has the property that by performing 
just {\em local} operations on one qubit, 
the complete Bell basis can 
be generated. That is,  states generated by local operations
are totally distinguishable. This remarkable property is due to maximal 
quantum entanglement between the two particles. We present a set of 
local unitary transformations that generate out of partially entangled
two-qubit state a set of four maximally distinguishable states
that are mutually equally distant. We discuss quantum dense coding based
on these alphabet states.
\end{abstract}

\begin{multicols}{2}

\section{Introduction}
Two parties (Alice \& Bob) who share a pure two-qubit state $|\Psi_1\r_{AB}$
can generate three other states $|\Psi_j\r_{AB}$ ($j=2,3,4$) such that the
four states form a basis in the Hilbert space of two qubits. In general,
the two parties have to perform operations 
on {\em both} qubits to generate the orthogonal states $|\Psi_j\r_{AB}$. 
Nevertheless, there is an exception - if the original state
$|\Psi_1\r_{AB}$ is one of the four Bell states \cite{Preskill98} then
by performing unitary transformations on just {\em one} of the two qubits
(let us assume Alice is the operations) the other three Bell states 
that form the Bell basis of the two-qubit system can be generated.
Specifically, let us assume the system is initially in the Bell state
\be
|\Psi_1\r_{AB}=\frac{1}{\sqrt{2}}\left(|0\r_{A}|0\r_{B} + 
|1\r_{A}|1\r_{B}\right),
\label{1.1}
\ee
where $|0\r_{X}$ and $|1\r_{X}$ ($X=A,B$) are basis vectors in the Hilbert
space ${\cal H}_X$ of the qubit $X$ (in what follows we will use the
shorthand notation $|00\r=|0\r|0\r$ and where clear 
we will omit subscripts indicating the subsystem). Now we introduce
four {\em local} (single-qubit)  operations 
\be
\hat{S}_1&=&\hat{\openone}~=~(|0\r\l0|+|1\r\l1|) ; 
\nonumber
\\
\hat{S}_2&=&\hat{\sigma}_x=~(|0\r\l1|+|1\r\l0|) ;
\nonumber
\\
\hat{S}_3&=&\hat{\sigma}_y=i(|0\r\l1|-|1\r\l0|) ;
\nonumber
\\
\hat{S}_4&=&\hat{\sigma}_z=~(|0\r\l0|-|1\r\l1|), 
\label{1.2}
\ee
where $\hat{\sigma}_\mu$ ($\mu=x,y,z$) are three Pauli operators.
When the operators $S_k$ act on the first (Alice's) qubit of 
the Bell state (\ref{1.1}) we find
\be
|\Psi_1\r &=&
\hat{S}_1\otimes\hat{\openone} |\Psi_1\r = \frac{1}{\sqrt{2}}
\left|00\r+|11\r\right) ;
 \nonumber
\\
|\Psi_2\r &=&
\hat{S}_2\otimes\hat{\openone} |\Psi_1\r = \frac{1}{\sqrt{2}}
\left|10\r+|01\r\right) ;
 \nonumber
\\
|\Psi_3\r &=&
\hat{S}_3\otimes\hat{\openone} |\Psi_1\r = \frac{i}{\sqrt{2}}
\left|01\r-|10\r\right) ;
 \nonumber
\\
|\Psi_4\r &=&
\hat{S}_4\otimes\hat{\openone} |\Psi_1\r = \frac{1}{\sqrt{2}}
\left|00\r-|11\r\right).
\label{1.3}
\ee
We see that the four states given by Eq.~(\ref{1.3}) are indeed four Bell
states \cite{Preskill98}. This means that by performing just {\em local}
operations the two-qubit states are changed globally in such a way that 
the four outcomes are perfectly distinguishable (i.e., the four Bell states
are mutually orthogonal). In fact, we can say that the four outcomes
are mutually equally (and maximally) distant, which can be expressed
by their mutual overlap ${\cal O}_{kl}$
\be
{\cal O}_{kl}=\left|\l\Psi_k|\Psi_l\r\right|^2=\delta_{kl}.
\label{1.4}
\ee
This remarkable property of Bell states is due to the quantum entanglement
between the two qubits \cite{Preskill98}. As suggested by Bennett and
Wiesner \cite{Bennett92}, this property can be utilized for the
{\em quantum dense coding}. The idea is as follows: Alice can perform
locally on her qubit four operations that result in four orthogonal
two-qubit states. So after she performs one of the 
possible operations
she sends her qubit to Bob. Then Bob can perform a  measurement
on the two qubits and determine with the fidelity equal to unity which
of the four operations has been performed by Alice. In this way, Alice
has transferred two 
bits of information via sending just a single two-level
particle. This theoretical scenario has been implemented experimentally
by the Innsbuck group \cite{Mattle96} using polarization entangled states
of photons. One can conclude that the 
entanglement in the case of two qubits 
can double a capacity of the quantum channel.

Recently, Barenco and Ekert \cite{Barenco95}, Hausladen et al.
\cite{Hausladen96}, and 
Bose et al. \cite{Bose00} have discussed how the channel capacity
depends on the degree of entanglement between the two qubits. Specifically,
these authors have analyzed the situation when initially Alice and Bob share
a two qubit system in a state
\be
|\psi_1\r= \alpha |00\r + \beta |11\r \, .
\label{1.5}
\ee
Then Alice is performing locally one of the four unitary operations
$\hat{S}_k$ given by Eq.(\ref{1.2}). As a result of these operations
  four possible
states $|\phi_k\r$ can be  generated: 
\be
|\phi_1\r &=&
\hat{S}_1\otimes\hat{\openone} |\psi_1\r = ~~~
\left(\alpha|00\r+\beta|11\r\right) ;
 \nonumber
\\
|\phi_2\r &=&
\hat{S}_2\otimes\hat{\openone} |\psi_1\r = ~~~
\left(\alpha|10\r+\beta|01\r\right) ;
 \nonumber
\\
|\phi_3\r &=&
\hat{S}_3\otimes\hat{\openone} |\psi_1\r = -i
\left(\alpha|10\r-\beta|01\r\right) ;
 \nonumber
\\
|\phi_4\r &=&
\hat{S}_4\otimes\hat{\openone} |\psi_1\r = ~~~
\left(\alpha|00\r-\beta |11\r\right).
\label{1.6}
\ee
These states represent the ``alphabet''
that is used in the given communication channel between Alice and Bob.
Not all of these alphabet states are mutually orthogonal. If we evaluate the
overlap ${\cal O}_{kl}$ we find
\be
{\cal O}_{kl} = 
|\langle\phi_{k}|\phi_{l}\rangle |^{2}=\left\{
\begin{array}{ll}
1 & {\rm if ~} k=l\\
\Delta^{2} & {\rm if~ } kl=12,21,34,43\\
0 & {\rm else}
\end{array}
\right.
\label{1.7}
\ee   
where $\Delta=|\alpha|^2 -|\beta|^2$. 
The fact that not all alphabet states
are mutually orthogonal leads to a decrease of the channel capacity that in
the present case is less than two. Nevertheless, for at least partially
entangled qubits the channel capacity 
is still larger than unity. 

As seen from Eq.(\ref{1.7}) the four states $|\phi_k\r$ are not
mutually equally distant. Some of them are mutually orthogonal,
but some of them have a nonzero overlap. The main goal of this 
paper is to find a set of {\em local} unitary operations
$\hat{U}_k$ that generate out of the state $|\psi_1\r$ given
by Eq.(\ref{1.5}) the alphabet $|\psi_k\r=\hat{U}_k|\psi_1\r$
with the elements that are equally distant, that is the mutual
overlaps of these four states are equal, and simultaneously we require
that they are as small as possible. In other words, the states are
mutually as distinguishable as possible. Formally, we are looking
for transformations $\hat{U}_k$ such that
\be
{\cal O}_{kl} = 
|\langle\psi_{1}|\hat{U}^\dagger_k \hat{U}_l
|\psi_{1}\rangle |^{2}=\left\{
\begin{array}{ll}
1 & {\rm for ~} k=l\\
{\cal O} & {\rm for~ } k\neq l
\end{array}
\right.
\label{1.8}
\ee   
with ${\cal O}$ being as small as possible. In addition, the transformations
under consideration have to fulfill the {\em Bell limit}, that is
for $\Delta\rightarrow 0$, when $|\psi_1\r\rightarrow|\Psi_1\r$,   
they have to generate four maximally entangled mutually orthogonal
two-qubit states. We note that this set of states 
is not necesarily equal to standard Bell states given by Eq.~(\ref{1.3}). 
In our case the explicit
form of these states is given by Eq.~(\ref{2.20}) with
$\alpha=\beta=1/\sqrt{2}$.

\section{Equally distant states}
\label{sec2}
It is well known that any pure bipartite state can be written in the Schmidt 
basis  \cite{Peres93} as given by Eq.(\ref{1.5}). In order to find the four
transformations $\hat{U}_k$ that fulfill the condition (\ref{1.8}) we
remind ourselves that the most 
general unitary transformation on a two-dimensional Hilbert space 
is an element from a four-parametric group ${\rm U}(2)$ 
\be
\label{2.1}
\hat{U}_{k}=e^{i\varphi_{k}}
\left[\cos\psi_{k}\hat{\openone}+i\sin\psi_{k}
(\vec{n}_{k}\hat{\vec {\sigma}})\right],
\ee
where $\vec{n}_{k}=(\sin\theta_{k}\cos\phi_{k},\sin\theta_{k}\sin\phi_{k},
\cos\theta_{k})$ is a  normalized vector around which the rotation is
performed by an angle $\psi_{k}$.

From the condition (\ref{1.8}) it follows that we have to solve the
following set of equations
\be
\label{2.2}
{\cal O}_{kl}
 &=&|\langle\psi_{1}|\hat{ W}_{kl}|\psi_{1}\rangle |^{2} 
\nonumber
\\
&=&\left|\ |\alpha |^{2} \l 0|\hat{W}_{kl}|0\r
+|\beta|^{2} \l 1|\hat{W}_{kl}|1\r
\right|^{2}
\nonumber
\\
&=& {\cal O} = {\rm  minimum}\, ,
\ee
where we have introduced a notation.
\be
\label{2.3}
\hat{W}_{kl}=\hat{U}^{\dagger}_{k}\hat{U}_{l}.
\ee 

Taking into account the relation 
between Pauli operators 
$\hat{\sigma}_{\mu}\hat{\sigma}_{\nu}=\delta_{\mu\nu}
\hat{\openone}- i\varepsilon_{\mu\nu\kappa}\hat{\sigma}_{\kappa}
$
and the relation
\be
(\vec{n}_k\cdot\hat{\vec{\sigma}})
(\vec{n}_l\cdot\hat{\vec{\sigma}})
=
\vec{n}_k \cdot \vec{n}_l  \hat{\openone} 
-i [\vec{n}_k\times \vec{n}_l]\cdot \hat{\vec{\sigma}} \, ,
\label{2.3a}
\ee
we can rewrite 
$\hat{W}_{kl}$ as
\be
\label{2.4}
\hat{W}_{kl}&=& \hat{\openone} \left(\cos\psi_{k}\cos\psi_{l}+\vec{n}_{k}
\cdot \vec{n}_{l}\sin\psi_{k}\sin\psi_{l}\right)
\nonumber
\\
&-& 
i\hat{\vec{\sigma}}\cdot 
[\vec{n}_{k}\times\vec{n}_{l}]\sin\psi_{k}\sin\psi_{l}
\\
&-&
i\hat{\vec{\sigma}}(
\sin\psi_{k}\cos\psi_{l} \ \vec{n}_{k}-\sin\psi_{l}\cos\psi_{k} \ 
\vec{n}_{l}).
\nonumber
\ee
Due to the fact that quantum 
 states are determined  up to global phase
we can omit phase
factors $\varphi_k$ in Eq.~(\ref{2.4}).

From Eq.~(\ref{2.2}) we see that only diagonal elements of the operators 
$\hat{W}_{kl}$ are relevant. Taking into account that 
only $\hat{\sigma}_{z}$ has nonvanishing diagonal elements, we obtain
\be
\label{2.5}
\l 0|\hat{ W}_{kl}|0\r &=& a_{kl}- i b_{kl} \, ,
\nonumber
\\
\l 1|\hat{ W}_{kl}|1\r &=& a_{kl}+i b_{kl} \, ,
\ee
where we use the notation
\be
\label{2.6}
a_{kl}&=&\cos\psi_{k}\cos\psi_{l}+\vec{n}_{k}\cdot\vec{n}_{l}
\sin\psi_{k} \sin\psi_{l} \, ,
\\
b_{kl}&=&\sin\psi_{k}\sin\psi_{l} [\vec{n}_{k}\times\vec{n}_{l}]_{z}
\nonumber
\\
&&+\sin\psi_{k}\cos\psi_{l}(\vec{n}_{k})_{z}
-\sin\psi_{l}\cos\psi_{k}(\vec{n}_{l})_{z}
\nonumber
\ee
and
\be
\label{2.7}
\vec{n}_{k} \cdot \vec{n}_{l}
 & = &
\cos\theta_{k}\cos\theta_{l}
+\sin\theta_{k}\sin\theta_{l}\cos (\phi_{l}-\phi_{k}) ; 
\nonumber
\\
\left[\vec{n}_k\times\vec{n}_l\right]_z 
 & = &
\sin\theta_{k}\sin\theta_{l}\sin (\phi_{l}-\phi_{k}).
\ee
It follows from Eq.~(\ref{2.2})  that 
\be
\label{2.8}
|\langle\psi_{k}|\psi_{l}\rangle|^{2}= {\cal O}_{kl}= 
a_{kl}^{2}+b_{kl}^{2}\Delta^{2}.
\ee
This overlap has to be minimized and made state-independent (i.e.,
${\cal O}_{kl}= {\cal O} = minimal$). 

In order to solve the problem, we 
choose $ \hat{U}_{1}=\hat{\openone}$ 
and explicitly rewrite the condition (\ref{2.8}) for $k=1$ and $l=2,3,4$:
\be
\label{2.9}
|\langle\psi_{1}|\psi_{l}\rangle|^{2}=
\cos^{2}\psi_{l}+\Delta^{2}\sin^{2}\psi_{l}\cos^{2}\theta_{l}.
\ee

From Eqs.~(\ref{2.8}) and (\ref{2.9}) it follows 
that in order 
to fulfill the Bell limit, when ${\cal O}=0$, two following conditions
have to be valid: 
\be
\label{2.10}
\cos^{2}\psi_{k}&=&0
\nonumber
 \\
\vec{n}_{k}\cdot\vec{n}_{l}&=&\delta_{kl}. 
\ee
Taking into account these constraints we  rewrite 
Eqs.(\ref{2.8}) and (\ref{2.9}) as
\be
\label{2.11a}
{\cal O}&=&
[\vec{n}_{k}\times\vec{n}_{l}]^{2}_{z}\Delta^{2} \, ,
\nonumber
\\
{\cal O} &=&\cos^{2}\theta_{k}\Delta^{2}, 
\ee
respectively.
Because the overlap ${\cal O}$ is supposed to be the same for all
pairs of states we can introduce a notation $F=\cos^2\theta_k$
and 
we  compare the right hand sides of Eqs.~(\ref{2.11a})
which gives us the following equation: 
\be
\label{2.11}
(1-F^{2})^{2}\sin^{2}(\phi_{k}-\phi_{l})-F^{2} = 0, 
\ee
where we have used the relation (\ref{2.7}).
From the condition $\vec{n_{k}}\cdot\vec{n_{l}}=0$ for
$k\neq l$ [see Eq.(\ref{2.10})] we write the constraint 
 for the parameter $F$
\be
\label{2.12}
(1-F^{2})\cos(\phi_{l}-\phi_{k})\pm F^{2}=0
\ee 
where $\mp=\rm Sgn(\cos\theta_{k}\cos\theta_{l})$.
The two constraints (\ref{2.11}) and (\ref{2.12}) are fulfilled when 
$F^{2}={1}/{3}$.

Now we  put our results together. We have found that  transformations
$\hat{U}_k$ 
are characterized by the following parameters: 
\be
\label{2.14}
\cos^{2}\psi_{k}&=&0 \, ,
\nonumber
\\ 
\cos^{2}\theta_{k}&=&\frac{1}{3} \, ,
\\
\cos(\phi_{l}-\phi_{k})&=&\frac{\mp F^{2}}{1-F^{2}}=\mp\frac{1}{2}\, . 
\nonumber
\ee
If we choose
$\cos\theta_{2}=\cos\theta_{3}={1}/{\sqrt{3}}$ 
then in order to have three distinct transformations we have to take
$\cos\theta_{4}=-{1}/{\sqrt{3}}$.
Our result (\ref{2.14}) also implies that 
\be
\label{2.17}
\cos(\phi_{2}-\phi_{3})&=&-1/2 \, ,
\\
\cos(\phi_{2}-\phi_{4})&=&\cos(\phi_{3}-\phi_{4})=1/2, 
\nonumber
\ee
which can be obtained when 
$\phi_{2}=\phi,\ \phi_{3}=\frac{2}{3}\pi+\phi ,\ 
\phi_{4}=\frac{\pi}{3}+\phi$. As we can see,  there is still
some freedom in a choice of the phase 
$\phi$. Just for convenience we take $\phi =0$. This finishes
our explicit construction of a set of {\em local} unitary transformations
$\hat{U}_k$ for which we have obtained the expressions
(here we have assumed that  in Eq.(\ref{2.1}) for 
the operators $\hat{U}_k$ the phase factors $\varphi_k$ are taken to
be equal to $\varphi_k=-\pi/2$)
\be
\label{2.18}
\hat{U}_1&=&\hat{\openone} \, ,
\\
\hat{U}_k&=&  \vec{n}_k \cdot \hat{\vec{\sigma}};\qquad k=2,3,4
\nonumber
\ee
where the unit 
vectors $\vec{n}_k$ are given by the expressions
\be
\label{2.19}
\vec{n}_{2}&=&\left(\frac{2}{\sqrt{6}}, 0, \frac{1}{\sqrt{3}}\right); 
\nonumber
\\
\vec{n}_{3}&=&\left(-\frac{1}{\sqrt{6}}, \frac{1}{\sqrt{2}},
\frac{1}{\sqrt{3}}\right) \, ;
\\
\nonumber
\vec{n}_{4}&=&\left(\frac{1}{\sqrt{6}}, \frac{1}{\sqrt{2}},
-\frac{1}{\sqrt{3}}\right) \, .
\ee
These vectors not only fulfill the condition $\vec{n}_k\cdot\vec{n}_l
=\delta_{kl}$ but also 
\be
\left[\vec{n}_k\times \vec{n}_l\right] = - \varepsilon_{klm}\vec{n}_m
\label{2.19a}
\ee
from which it follows that we can rewrite the operator $\hat{W}_{kl}
=\hat{U}^\dagger_k\hat{U}_l$ for $k,l=2,3,4$ as
\be
\hat{W}_{kl}=\delta_{kl}\hat{\openone} + \varepsilon_{klm}\hat{U}_m
\label{2.19b}
\ee

The operators $\hat{U}_k$ generate from the reference state
$|\psi_1\r$ via local transformations 
the alphabet with most distant states. We stress that for a chosen
Schmidt basis the local operators $\hat{U}_k$ do not depend on the
states to be rotated. In this sense, these operators are {\em universal}.
The alphabet states  
in the basis $\{\ |00\r,|10\r,|01\r,|11\r \}$ read 
\be
\label{2.20}
|\psi_{1}\rangle &=&(\alpha ,0,0,\beta )\, ;
\nonumber
\\ 
|\psi_{2}\rangle&=&\left(\frac{1}{\sqrt 3}\alpha ,\sqrt\frac{2}{3}\alpha ,
\sqrt\frac{2}{3}\beta ,-\frac{1}{\sqrt 3}\beta\right) \, ;
\\ 
|\psi_{3}\rangle
&=&\left(\frac{1}{\sqrt 3}\alpha ,\frac{-1-i\sqrt{3}}{\sqrt 6}\alpha ,
\frac{-1+i\sqrt{3}}{\sqrt 6}\beta ,-\frac{1}{\sqrt 3}\beta\right) \, ;
\nonumber
\\ 
|\psi_{4}\rangle &=&\left(-\frac{1}{\sqrt 3}\alpha
,\frac{1-i\sqrt{3}}{\sqrt 6}\alpha ,\frac{1+i\sqrt{3}}{\sqrt 6}\beta
,\frac{1}{\sqrt 3}\beta\right)\, .
\nonumber
\ee   
By construction, the mutual overlap between these states is minimal and
equal to
\be
\label{2.21}
{\cal O}= \frac{1}{3} \Delta^2.
\ee

{\em Comment 1.}\newline
We note that the universal transformations we have derived generate
a set of four states $|\psi_k\r$ for the state $|\psi_1\r$. In fact,
these transformations generate the same set of states if generated
from any state from this set (that is, we observe a specific permutation
invariance in the set). To prove this property it is enough to observe that
\be
\hat{U}_{k}\hat{U}_l=
\delta_{kl}\hat{\openone} + \varepsilon_{klm}\hat{U}_m \, ,
\label{2.22}
\ee
which means that the state $|\phi_k\r=\hat{U}_k|\psi_l\r$  is equal
to one of the states $|\psi_m\r=\hat{U}_m |\psi_1\r$ given by
Eqs.(\ref{2.20}).

{\em Comment 2.}\newline
We have derived our transformations under the assumption that the reference
state from which the other three alphabet states are generated is a pure
state. When the reference state $\hat{\rho}_1$ is a statistical mixture
of two qubits, which in general is characterized by 
15 parameters, our transformations generate an alphabet $\hat{\rho}_k
= \hat{U}_k\hat{\rho}_1\hat{U}_k^\dagger$ 
such that in general 
${\rm Tr}(\hat{\rho}_k\hat{\rho}_l)\neq const$. Nevertheless, for a large
class of statistical mixtures of two qubits the transformations $\hat{U}_k$
generate equally distant alphabets. A simple example would be to consider
the reference state to be a mixture of the form
$\hat{\rho}_1=s|\psi_1\r\l\psi_1| +\frac{1-s}{4}\hat{\openone}$. In 
this case the overlap between alphabet states is constant and equal to
${\cal O}={\rm Tr}(\hat{\rho}_k\hat{\rho}_l) = s^2\Delta^2/3
+(1-s^2)/4$.  Another example is when the reference state is taken to be
$\hat{\rho}_1=\sum_m \lambda_m\hat{U}_m|\psi_1\r\l\psi_1|\hat{U}_m^\dagger$
and the three alphabet states are generated by the operators $\hat{U}_k$.
In this case we find that the overlap between the states is constant and
equal to ${\cal O}=\frac{1}{3}\Delta^2 \sum_{m,n=1}^4 \lambda_m\lambda_n$.

\section{Channel capacity}
Let us assume that Alice and Bob are using alphabet states
described above for quantum communication. The capacity of
the quantum channel is given by the expression \cite{Holevo98}
\begin{equation}
\label{holevo}
C=\max_{\pi}\left[ 
S(\sum\pi_{k}\hat{\varrho}_{k})-\sum\pi_{k}S(\hat{\varrho}_{k})\right],
\end{equation}
where $\hat{\varrho}_{k}$ are the alphabet states at the output
of the channel (i.e., at Bob's side of the communication channel) 
and $\pi_k$ are the 
probabilities with which the alphabet states are used by Alice.
In the right-hand side of Eq.~(\ref{holevo}), the function
$S$ is the von Neumann entropy $S(\hat{\varrho})
=-{\rm Tr}(\hat{\varrho}\log_2 \hat{\varrho})$.

Firstly, we analyze ideal channels and then we describe quantum
capacity of noisy channels.

\subsection{Ideal channel}
In the case of the ideal channel we evaluate 
 the capacity for two sets of alphabet states - the
one used by Bose et al. given by Eq.~(\ref{1.6}) - 
and the other set we have derived earlier in the paper 
[see Eq.~(\ref{2.20}]. We denote the reference pure state
from which the elements of 
alphabets are generated as $\hat{\varrho}_{AB}=|\psi_1\rangle
\langle \psi_1 |$, where $|\psi_1\rangle$ is given by Eq.~(\ref{1.5}).

Interestingly enough, for both cases we find
the capacity of the quantum channel to be the same: 
\begin{eqnarray}
C=1+S(\hat{\varrho}_{A}),
\label{3.2}
\end{eqnarray}
where $\hat{\varrho}_A ={\rm Tr}_B \hat{\varrho}_{AB}$.
Obviously, in the Bell limit, when the $|\psi_1\rangle$ is
equal to a Bell state and  Alice's qubit is 
in a maximally mixed state with  $S(\hat{\varrho}_A)=1$,
the capacity of the quantum channel is equal to 2.

But the question is: Why for the two alphabets discussed above
is the quantum capacity is mutually {\em equal} for an arbitrary
reference state $|\psi_1\rangle$? To illuminate this problem
we remind ourselves 
that the two sets of the operators that  generate
two alphabets (\ref{1.6}) and (\ref{2.20}), respectively,
fullfil the ``Bell's limit'', i.e.
$\vec{n}_k.\vec{n}_l =\delta_{kl}$ [see Eq.~(\ref{2.10})].
Therefore we concentrate our attention on those transformations
 (\ref{2.18})
that  have this property.

Because of the unitarity of these transformations the second term 
in Eq.~(\ref{holevo}) is equal to zero.
The input probability 
that maximize the expression for capacity is equal to
$\pi_k =1/4$. In this case the density operator 
$\hat{\overline{\varrho}}=\sum_k\hat{\varrho}_k/4$
 in the matrix form (in the basis $\{|00\rangle,
|10\rangle, |01\rangle,|11\rangle\}$)
reads
\end{multicols}
\vspace{-0.5cm}
\noindent\rule{0.5\textwidth}{0.4pt}\rule{0.4pt}{\baselineskip}
\widetext 
\begin{eqnarray}
\hat{\overline{\varrho}}=\frac{1}{4}\left(
\begin{array}{cccc}
|\alpha |^{2}(1+\sum_{k}n_{k}^{z} n_{k}^{z}) 
& |\alpha |^{2}\sum_{k}n_{k}^{z}(n_{k}^{x}-in_{k}^{y}) 
& \alpha\beta^{*}\sum_{k}n_{k}^{z}(n_{k}^{x}+in_{k}^{y}) 
&  \alpha\beta^{*}(1-\sum_{k}n_{k}^{z} n_{k}^{z})\\
 |\alpha |^{2}\sum_{k}n_{k}^{z}(n_{k}^{x}+in_{k}^{y}) 
& |\alpha |^{2}\sum_{k}(n_{k}^{y} n_{k}^{y}+n_{k}^{x} n_{k}^{x}) 
&  \alpha\beta^{*}\sum_{k}(n_{k}^{x}+in_{k}^{y})(n_{k}^{x}+in_{k}^{y}) 
&   \alpha\beta^{*}\sum_{k}n_{k}^{z}(n_{k}^{x}-in_{k}^{y})\\
\alpha^{*}\beta\sum_{k}n_{k}^{z}(n_{k}^{x}-in_{k}^{y}) 
& \alpha^{*}\beta\sum_{k}(n_{k}^{x}-in_{k}^{y})(n_{k}^{x}-in_{k}^{y}) 
& |\beta |^{2}\sum_{k}(n_{k}^{y} n_{k}^{y}+n_{k}^{x} n_{k}^{x}) 
&  |\beta |^{2}\sum_{k}n_{k}^{z}(n_{k}^{x}+in_{k}^{y})\\
\alpha^{*}\beta (1-\sum_{k}n_{k}^{z} n_{k}^{z}) 
&  \alpha^{*}\beta\sum_{k}n_{k}^{z}(n_{k}^{x}+n_{k}^{y}) 
&  |\beta |^{2}\sum_{k}n_{k}^{z}(n_{k}^{x}-in_{k}^{y}) 
& |\beta |^{2}(1+\sum_{k}n_{k}^{z} n_{k}^{z})\\
\end{array}
\right)
\end{eqnarray}
\begin{multicols}{2}
where  $n_{k}^{j}$ denotes the j-th component of the vector $\vec{n}_{k}$. 
These three-dimensional vectors create a complete system in 
three-dimensional real vector space, i.e., 
\begin{eqnarray}
\sum_{k=2}^4 n_{k}^{j}n_{k}^{l}=\delta^{jl},
\end{eqnarray}
for $j,l=x,y,z$.
Using this property, we  evaluate the operator 
$\hat{\overline{\varrho}}$ for which we find
\begin{equation}
\label{ro}
\hat{\overline{\varrho}} =
\frac{|\alpha |^{2}}{2}(|00\rangle\langle 00|+|10\rangle\langle 10|)+
\frac{|\beta |^{2}}{2} (|01\rangle\langle 01|+|11\rangle\langle 11|).
\end{equation} 
The corresponding quantum capacity
of the ideal channel with pure signal states then reads
\begin{equation}
\label{vyslednakapacita}
C=1-|\alpha |^2\log{|\alpha |^2}-|\beta |^2\log{|\beta
|^2}=1+S(\hat{\varrho}_A).
\end{equation}
and is equal for {\em all} alphabets which satisfy the condition
(\ref{2.10}).

{\em Comment 3}\newline
The capacity (\ref{vyslednakapacita}) is the biggest
possible capacity of the quantum channel for alphabets
that are generated by local operations from the reference
state (\ref{1.5}). To see this, we can
imagine for a while that 
 there exist four local unitary 
transformations that generate the alphabet for 
which the capacity is bigger than 
(\ref{vyslednakapacita}). In the case of a maximally 
entangled state they must fulfill the Bell limit, 
i.e., the alphabet is an 
 orthogonal basis. On the other hand, we have shown that all 
 transformations that satisfy the Bell limit have to fulfill
the condition (\ref{2.10}). Consequently, they have to belong
to the set of our equivalent transformations, with the
channel capacity (\ref{vyslednakapacita}). This contradicts
the original assumption, which proves our statement.

\subsubsection{Mixed reference state}
Let us assume that the reference state $\hat{\varrho}_{AB}$ shared by
Alice and Bob is a statistical mixture that is parameterized as
\begin{equation}
\label{zmes}
\hat{\varrho}_{AB}=\sum_{j=1}^{4}
\lambda_{j}|\chi_{j}\rangle\langle\chi_{j}|.
\end{equation}
In this spectral decomposition the orthogonal states 
$|\chi_j \rangle$ can be written in 
the same Schmidt basis for all $j=1,2,3,4$:
\begin{eqnarray}
\begin{array}{l}
|\chi_{1}\rangle =
\alpha |0\rangle_{A}|0\rangle_{B}+\beta |1\rangle_{A}|1\rangle_{B}\\
|\chi_{2}\rangle =
\beta^{*}|0\rangle_{A}|0\rangle_{B}-\alpha^{*}|1\rangle_{A}|1\rangle_{B}\\ 
|\chi_{3}\rangle =
\gamma |0\rangle_{A}|1\rangle_{B}+\delta|1\rangle_{A}|0\rangle_{B}\\
|\chi_{4}\rangle =
\delta^{*}|0\rangle_{A}|1\rangle_{B}-\gamma^{*}|1\rangle_{A}|0\rangle_{B}.
\end{array}
\end{eqnarray}
(note that $|\chi_1\rangle=|\psi_1\rangle$).

With this reference state 
the alphabet is the set of states
$\hat{\varrho}_{k}=\hat{U}_{k}\hat{\varrho}_{AB}
\hat{U}_{k}^{\dagger}$ generated by
the set of four local transformations 
\{$\hat{U}_1 =\hat{\openone} ,
\hat{U}_k =\vec{n}_k.\vec{\sigma}$\}, as before.

In this case the second term in 
the expression (\ref{holevo}) for channel capacity 
 does not vanish. 
Our transformations are unitary. Therefore 
the entropy for state $\hat{\varrho}_k$ is the same and equals to 
\begin{eqnarray}
S(\hat{\varrho}_{AB})=-\sum_{j}\lambda_{j}\log\lambda_{j}.
\end{eqnarray}

To evaluate the final expression for the channel capacity we
have to find 
the entropy of the state
\begin{eqnarray}
\hat{\overline{\varrho}}
=\frac{1}{4}\sum_{k=1}^{4}\hat{U}_{k}
\hat{\varrho}_{AB}\hat{U}_{k}^{\dagger}=
\sum_{j=1}^{4}\lambda_{j}
\frac{1}{4}\sum_{k=1}^{4}
\hat{U}_{k}|\chi_{j}\rangle\langle\chi_{j}|\hat{U}_{k}^{\dagger}.
\end{eqnarray}
The term $\frac{1}{4}\sum_{k=1}^{4}\hat{U}_{k}|\chi_{j}
\rangle\langle\chi_{j}|\hat{U}_{k}^{\dagger}$ 
is for all $j=1,2,3,4,$ diagonal as in Eq.~(\ref{ro}). 
It means that $\hat{\overline{\varrho}}$ is diagonal in the 
given Schmidt basis
\begin{eqnarray}
\hat{\overline{\varrho}} =\frac{x}{2}(|00\rangle\langle 00|
+|10\rangle\langle 10|)+
\frac{y}{2}(|01\rangle\langle 01| +|11\rangle\langle 11|).
\end{eqnarray}
where
\begin{eqnarray}
x&=&|\alpha |^{2}\lambda_{1}+|\beta |^2 \lambda_{2}+
|\gamma |^{2}\lambda_{3}+|\delta|^2 \lambda_{4} \, , 
\nonumber
\\
\label{xy}
y&=&|\alpha |^{2}\lambda_{2}+|\beta |^2 \lambda_{1}+
|\gamma |^{2}\lambda_{4}+|\delta |^2 \lambda_{3}.
\end{eqnarray}
Finally, 
taking into account that the reduced density operator $\hat{\varrho}_A$
has the form
\begin{eqnarray}
\hat{\varrho}_{A}=
{\rm Tr}_{B}(\hat{\varrho}_{AB})=\left(
\begin{array}{cc}
x & 0 \\
0 & y \\
\end{array}
\right) \, ,
\end{eqnarray}
we can express the 
 capacity of the ideal channel as 
\begin{eqnarray}
\label{kapacitaprezmesi}
C&=&\sum_{j}\lambda_{j}\log\lambda_{j}+1-x\log{x}-y\log{y}
\nonumber
\\
&=&1+S(\hat{\varrho}_{A})-S(\hat{\varrho}_{AB}).
\end{eqnarray}

\subsection{Pauli channel}
From above, it follows 
alphabets that fulfill the condition (\ref{2.10}) 
lead to  the same capacity of the {\em ideal} quantum channel.
Let us assume that the channel is noisy. We will model
an imperfect channel as a Pauli channel 
 \cite{Preskill98} characterized by the parameters
$p_x, p_y, p_z$ and $p=p_x+p_y+p_z$.
In this case, the alphabet states that are used
for coding at the output 
can be expressed as
\begin{equation}
\hat{\rho}_k' =
(1-p)|\psi_k \rangle \langle \psi_k | + \sum_{\mu=x,y,z}
p_\mu \hat{\sigma}_\mu|\psi_k\rangle\langle\psi_k| 
\hat{\sigma}_\mu,
\label{3.1}
\end{equation}
(here we implicitly assume that
Bob's qubit is left intact).
Taking into account the
explicit expression for the operators $\hat{U}_k$,  we find
\begin{equation}
\hat{\sigma}_\mu\hat{U}_k = \sum_{\nu=x,y,z} n_k^{(\nu)} \left[
\delta_{\mu\nu}
\hat{\openone}-i\varepsilon_{\mu\nu\kappa}\hat{\sigma}_\kappa 
\right]\, .
\label{3.2a}
\end{equation}
With the help of the last expression, we rewrite  the density
operator (\ref{3.1}) as 
\begin{eqnarray}
\label{3.12}
\hat{\varrho}_{k}'&=&(1-p)\sum_{\mu ,\nu
=1}^{3}\hat{\sigma}_{\mu}\hat{\varrho}_{0}
\hat{\sigma}_{\nu}n_{k}^{\mu}n_{k}^{\nu}
\nonumber
\\
&+&\sum_{\mu =1}^{3}p_{\mu}\{ (n_{k}^{\mu})^{2}\hat{\varrho}_{0}
+i n_{k}^{\mu}[(\vec{n}_{\mu}\times\vec{n}_{k})\cdot \hat{\vec{\sigma}},
\hat{\varrho}_{0}]
\\
&+&[(\vec{n}_{\mu}\times\vec{n}_{k})\cdot \hat{\vec{\sigma}}]
\hat{\varrho}_{0}[(\vec{n}_{\mu}\times\vec{n}_{k})\cdot \hat{\vec{\sigma}}]\}
\nonumber
\end{eqnarray}
where $\hat{\varrho}_{0}$ denotes the reference state from which
the alphabet is generated,
and $\vec{n}_\mu $ is the vector 
defined by $\vec{n}_\mu\cdot \hat{\vec{\sigma}}=
\hat{\sigma}_\mu $ for $\mu =x,y,z$.
So this specifies the alphabet used. Now we want to evaluate the
capacity of the channel (\ref{holevo}). 
We assume the input probability
$\pi_k=1/4$ and in this case 
\begin{eqnarray}
\hat{\overline{\varrho}}
=\frac{1}{4}(\hat{\varrho}_{0}+\sum_{\mu}
\hat{\sigma}_{\mu}\hat{\varrho}_{0}\hat{\sigma}_{\mu}),
\label{3.13} 
\end{eqnarray}
which  is the same as Eq.~(\ref{ro})!
So only the second term in Eq.~(\ref{holevo}) can be 
different for different choice of Alice transformations.

Let us assume that Alice and Bob are using for  communication 
two  alphabets (\ref{2.20}) and (\ref{1.6}), respectively.
We want to find which alphabet gives us a higher capacity of
an imperefect Pauli channel.

\subsubsection{Depolarizing channel}
\begin{figure}
\centerline {\epsfig{width=7.0cm,file=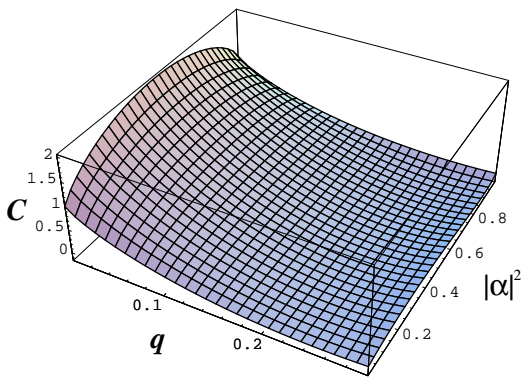}}
\begin{narrowtext}
\bigskip      
\caption{ 
We plot the capacity of the depolarizing channel as a function of the
parameters $q$ and $|\alpha|^2$ that characterize the alphabet used. 
 }
\label{fig1}
\end{narrowtext}
\end{figure}

Firstly, let us assume a
depolarizing channel, for which $p_x=p_y=p_z\equiv q$ with $0\leq q \leq 1/3$.
In this case  
the operators $\hat{\rho}_k'$ given by Eq.(\ref{3.1}) 
have the same
eigenvalues for both alphabets that read
\begin{eqnarray}
\label{3.5}
\eta_{1} &=&2  q|\alpha |^{2}; \qquad 
\eta_{2}=2 q|\beta |^{2} \, ;
\\
\nonumber
\eta_{3}&=&\frac{1}{2}(1-2 q
+\sqrt{(1-2 q)^{2}-16 q|\alpha |^{2}|\beta |^{2}(1-3 q)}) \, ;
\\
\nonumber
\eta_{4}&=&\frac{1}{2}(1-2 q
-\sqrt{(1-2 q)^{2}-16 q|\alpha |^{2}|\beta |^{2}(1-3 q)})\,  ,
\end{eqnarray}
so that the capacity can be expressed as 
$C=S(\hat{\overline{\rho}})-S(\hat{\rho}_k' )$.
We plot this capacity in Fig.~\ref{fig1}. 
We clearly see that the larger
the degree of entanglement  the greater is the capacity of the quantum 
channel, 
irrespectively, on the value of the parameter $q$.

From above it follows that for the depolarizing channel both alphabets
provide us with the same capacity. 

\subsubsection{x-Pauli channel}
Now
our task is to present an example that illustrates that with the 
equally distant alphabet (\ref{2.10}) Alice can perform better
(i.e.,  the channel capacity is higher) than with the standard alphabet
(\ref{1.6}). Let us assume the channel such that 
 $p_y =p_z =0$ with $0\geq p_x\geq 1$. In this case we find
two nonzero eigenvalues of the output state
$\hat{\varrho}_k' $ (\ref{3.12})
\begin{eqnarray}
\eta_{\pm}=\frac{1}{2}(1\pm 4p(1-p_x )\delta_k^2 ), 
\end{eqnarray}
where 
$\delta_k^2 =1-|\langle\psi_k |\hat{\sigma}_x |\psi_k \rangle |^2$, 
$|\psi_k\rangle =\hat{U}_k\otimes\hat{\openone} |\psi_0\rangle$
with 
\begin{eqnarray}
\langle\psi_k |\hat{\sigma}_x |\psi_k \rangle =
2n_k^z n_k^x (\alpha^2 -\beta^2).
\end{eqnarray}
For the standard alphabet (\ref{1.6})
we find  $\delta_k =1$ for all $k$, while for
the equally distant alphabet (\ref{2.10}) 
$\delta_1 =1,\delta_2^2 =1-8\triangle^2 /9,
\delta_3^2 =\delta_4^2 =1-2\triangle^2 /9$.

\begin{figure}
\centerline {\epsfig{width=7.5cm,file=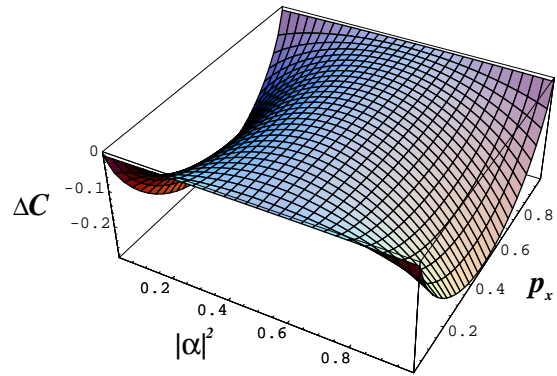}}
\begin{narrowtext}
\bigskip      
\caption{ 
We plot  the difference 
between the capacity using the standard alphabet (\ref{1.6})
and the equally distant alphabet. This difference is plotted
as a function of 
 $|\alpha |^2$
and $p_x$ that characterizes the x-Pauli channel. }
\label{fig2}
\end{narrowtext}
\end{figure}

Using these results, we  directly evaluate the two capacities
of our interest. We note that for both alphabets the operator 
$\hat{\overline{\varrho}}$ is the same [see Eq.(\ref{3.13})].
 Consequently, the entropy $S(\hat{\overline{\varrho}})$ in
the expression for the channel capacity is the same.
Therefore, the only difference can arise from the
terms $S(\hat{\varrho}_k ')$. This entropy is determined
by the eigenvalues $\eta_\pm$. Obviously, 
 the closer  the eigenvalues are  to 1/2 the larger  
is the entropy $S(\hat{\varrho}_k' )$ and the smaller
is the capacity. We see that in the case of the
standard alphabet the eigenvalues are closer to 1/2 than
in the case of the equally distant alphabet. Therefore
we conclude that the second alphabet leads to a higher
channel capacity for the given  x-Pauli channel. We plot the 
difference between the standard and the equally-distant
alphabet capacities in Fig.~\ref{fig2}.

\section{Conclusions}

In this paper, we have presented a set of four local unitary operators that 
generate from a partially entangled pure two-qubit state a set of 
equally distant states with a minimal overlap. We have
evaluated capacity of an ideal and Pauli channels using this
alphabet. We have shown that in some cases our alphabet leads 
to a higher channel capacity than the standard alphabet used
by Bose et al. \cite{Bose00}.

We conclude that in order to validate the capacity of our quantum
channel Alice has to use a block coding scheme for sending a message.
Bob on his end has to perform a collective measurement on the whole
message rather than individual letters (alphabet states).The explicit
expression for this collective decision rule is given in
Ref.~\cite{Holevo98}. 
As shown by Holevo \cite{Holevo98} and Hausladen et al. \cite{Hausladen96}, 
in this case  the information transmitted per letter can be made
arbitrarily close to the channel capacity (\ref{holevo}).

\acknowledgements
This work was supported 
by the IST project EQUIP under the contract
IST-1999-11053 and by  the CREST, Research Team
for Interacting Career Electronics. We  thank 
Sougato Bose for discussions.

\end{multicols}
\end{document}